\documentclass[12pt,preprint]{aastex}

\newcommand{\pks}{PKS~2155--304}
\newcommand{\rmax}{$r_{\rm max}$}
\newcommand{\tpeak}{$\tau_{\rm peak}$}
\newcommand{\tcent}{$\tau_{\rm cent}$}

\newcommand{\fvar}{$F_{\rm var}$}
\newcommand{\fpp}{$F_{\rm pp}$}

\newcommand{\xmm}{{\it XMM-Newton}}
\newcommand{\sax}{{\it BeppoSAX}}
\newcommand{\asca}{{\it ASCA}}

\newcommand{\et}{et al.\ }
\newcommand{\A}{$\rm \AA$}

\slugcomment{Accepted by ApJ}

\shortauthors{Zhang et al. }
\shorttitle{Multiwavelength Variability of PKS~2155--304}

\begin{document}

\title{Multiwavelength Observations of the BL Lacertae Object PKS~2155--304 with XMM-Newton}

\authoraddr{ }

\author{Y.H. Zhang\altaffilmark{1}, 
        J.M. Bai\altaffilmark{2},    
	S.N. Zhang\altaffilmark{1},    
        A. Treves\altaffilmark{3},  
        L. Maraschi\altaffilmark{4},  and
	A. Celotti\altaffilmark{5}
}

\altaffiltext{1}{Department of Physics and Tsinghua Center for Astrophysics (THCA), Tsinghua University, Beijing 100084, China; youhong.zhang@mail.tsinghua.edu.cn}
\altaffiltext{2}{National Astronomical Observatories/Yunnan Observatory, Chinese Academy of Sciences, P. O. Box 110, Kunming, Yunnan, 650011, China}
\altaffiltext{3}{Dipartimento di Scienze, Universit\`a degli Studi dell'Insubria, via Valleggio 11, I-22100 Como, Italy}
\altaffiltext{4}{Osservatorio Astronomico di Brera, via Brera 28, I-20121 Milano, Italy}
\altaffiltext{5}{International School for Advanced Studies, SISSA/ISAS, via Beirut 2-4, I-34014 Trieste, Italy}

\begin{abstract}

The optical-UV and X-ray instruments on-board \xmm\ provide an excellent opportunity to perform simultaneous observations of violently variable objects over a broad wavelength range. The UV and X-ray bright BL Lac object PKS~2155--304 has been repeatedly observed with \xmm\ about twice per year. In this paper, we present a detailed analysis of the simultaneous multiwavelength variability of the source from optical to X-rays, based on the currently available \xmm\ observations. These observations probed the intra-day multiwavelength variability at optical-UV and X-ray wavelengths of the source. The UV variability amplitude is substantially smaller than the X-ray one, and the hardness ratios of the UV to X-rays correlates with the X-ray fluxes: the brighter the source, the flatter the UV-X-ray spectra. On 2000 May 30-31 the UV and X-ray light curves were weakly correlated, while the UV variations followed the X-ray ones with no detectable lags on 2000 November 19-21. On 2001 November 30 the source exhibited a major X-ray flare that was not detected in the optical. The intra-day UV and X-ray variability presented here is not similar to the inter-day UV and X-ray variability obtained from the previous coordinated extensive multiwavelength campaigns on the source, indicating that different ``modes'' of variability might be operating in \pks\ on different timescales or from epoch to epoch. 

\end{abstract}



\section{Introduction}\label{sec:intro}

Rapid and large amplitude variability has been frequently detected on different timescales across the whole accessible electromagnetic spectrum of blazars (see Ulrich, Maraschi \& Urry 1997 for a review). The non-thermal continuum emission  is almost certainly produced by relativistic particles in a tangled magnetic field, in a relativistic jet closely aligned with the line of sight (e.g., Urry \& Padovani 1995). Relativistic beaming therefore plays a crucial role in the observed properties of blazars.

Multiwavelength observations have established that the spectral energy distributions (SEDs) of blazars consist of two rather smooth components in the $\nu F_{\nu}$ representation. The SED peak energies are most likely luminosity-related: the lower the luminosity, the higher the peak energy (e.g., Fossati \et 1998). For low-luminosity BL Lac objects like the well-studied sources Mrk~421, Mrk~501 and \pks, the low energy component typically peaks at UV-soft X-ray bands and for this reason these sources are usually called high-energy peaked BL Lac objects (HBLs). The primary mechanism for this component is believed to be synchrotron radiation of relativistic electrons. The high energy component is relatively poorly understood. For HBLs that have been detected at TeV energies, the high energy component appears to peak in the TeV band and its origin is probably inverse Compton scattering off soft photons, presumably synchrotron photons within the jet, as envisaged by the synchrotron self-Compton (SSC) process (e.g., Urry \& Padovani 1995). However, it is still unknown how the emitting particles are accelerated to energies high enough to produce the observed emission, and how the plasma is accelerated to a relativistic bulk velocity.  

HBLs are highly variable in the X-rays, as expected from the interpretation that the X-ray emission corresponds to the high energy tail of the synchrotron component. X-ray observations over the last decade with various X-ray telescopes have revealed a remarkably complicated behaviour of the X-ray variability (see Pian 2002 and Zhang 2003 for reviews), which can not be interpreted by simple models. Flux variations are generally accompanied by spectral changes. The X-ray spectrum typically hardens when the source is brighter. Systematic spectral evolution  has been resolved for well-defined flares: the peak of the synchrotron emission shifts to higher energy at higher flux level (e.g., Fossati \et 2000; Massaro \et 2004b; Tanihata \et 2004). Exceptionally high peak energies of the low energy component -- up to $\sim 100$~keV -- were detected in Mrk~501 (Pian \et 1997; Sambruna \et 2000). 

The correlations and possible time lags of variability in different energy bands are also remarkably complicated. The emission at low X-ray energies generally correlates with that at high energies. However, the maximum of the cross-correlation function (CCF) can occur at substantially different time lags (e.g., Tanihata \et 2001 for Mrk~501; Zhang \et 1999 for \pks; Ravasio \et 2004 for Mrk~421). Statistically weaker correlations were also found in several other cases. Lags of about two hours derived from the X-ray data gathered with low Earth orbit satellites (ASCA and \sax) were questioned by the earlier results inferred from \xmm\ data, mainly on the basis of the fact that the continuous observations of \xmm\ have relatively higher signal-to-noise ratio and temporal resolution than periodically interrupted observations of \asca\ and \sax\ do (Edelson \et 2001; Sembay \et 2002). This claim was disputed by Zhang \et (2004) who performed detailed Monte Carlo simulations. The results obtained with \sax\ and \asca\ have been recently supported by the finding of lags of $\sim$~1000~s and $\sim$~one hour in the \xmm\ observations of Mrk~421 and \pks\ (Ravasio \et 2004; Zhang \et 2006).

Recently, using \xmm\ timing mode observations of Mrk~421, Brinkmann \et (2005) presented a time resolved cross-correlation analysis between the soft and hard energy bands. They showed that the correlations, with different lags of both signs, change on a characteristic timescale of a few $10^{3}$~s, in qualitative agreement with previous claims about changes of lags during individual flares (e.g., Zhang \et 2002). Furthermore, it appears that the lag may be related to the flare duration: the longer the flare duration, the larger the lag (Zhang \et 2002; Brinkmann \et 2003), and to the spectral slope: a steeper slope corresponds to a higher probability that the soft band emission leads the hard one (Zhang \et 2006). The lags generally become larger, and the correlation weaker, with increasing differences of the compared energy bands (e.g., Zhang \et 2006; Ravasio \et 2004). Zhang (2002) also found preliminary evidence that in Mrk~421 the lags become larger with increasing timescale, as typically occurs in X-ray binaries.   

By measuring variability correlations, time lags and spectral variations, simultaneous multiwavelength observations offer the strongest constraints on blazar models, namely on the processes of particles injection/acceleration and diffusion in the emission region(s), and on the relative location of regions dominating the emission at different energies. The previous coordinated multiwavelength observations of a few bright blazars have revealed quite different variability patterns. For example, the variability around the peaks of the two SED components was well correlated in Mrk~421 (Maraschi \et 1999), while in \pks\ the fluxes at different wavelengths were not correlated on the measured timescales (Aharonian \et 2005b). It is important to probe whether the complex X-ray variability patterns mentioned above can be directly extrapolated to the UV and optical ranges since the optical-UV-X-ray emissions have a common (synchrotron) origin for HBLs. This diagnostics has to be carried out through simultaneous observations. Unlike other coordinated multiwavelength campaigns that need different telescopes at different sites, the optical monitor (OM) and X-ray telescopes onboard \xmm\ provide a very rare opportunity to simultaneously monitor variable sources like blazars from optical to X-rays. \xmm\ is thus particularly apt to study the high energy part of the synchrotron component of HBLs. 

The IUE and EUVE satellites revealed that \pks\ is the brightest BL Lac object at ultraviolet wavelength (Pian \et 1997; Marshall \et 2001 and reference therein). This is consistent with the possibility that the synchrotron emission of \pks\ peaks at EUV or even UV bands, as indicated by the fact that multi-epoch X-ray spectral analysis of \sax\ observations did not reliably find spectral peaks in such band (Zhang \et 2002). In contrast, in the other two well studied HBLs Mrk~421 and Mrk~501, the synchrotron emission peaks in high energy X-ray band (Pian \et 1998; Fossati \et 2000; Tanihata \et 2004; Massaro \et 2004a). \pks\ is also very bright in the X-rays (e.g., Zhang \et 2002), optical (e.g., Zhang \& Xie 1996 and references therein) and other lower energy bands dominated by synchrotron emission (e.g., Pesce \et 1997). \pks\ has been also detected in the 30~MeV--10~GeV band by EGRET (Vestrand, Stacy, \& Sreekumar 1995), showing a hard spectrum consistent with an SSC component. This indicates that \pks\ is a good candidate source of TeV emission. Indeed, the Durham Mk VI telescopes firstly detected its very high energy $\gamma$-ray emission (Chadwick \et 1999), recently confirmed by H.E.S.S at the $45\sigma$ significance level and still detectable even during the dark periods of the H.E.S.S observations (Aharonian \et 2005a). The brightness of \pks\ through such a broad energy range makes it one of the most suited blazars for carrying out multiwavelength observations to test emission mechanisms in detail (e.g., Urry \et 1997; Mimica \et 2005), and for possibly distinguishing between leptonic vs hardronic origins of the emission (Aharonian \et 2005b). The intense monitorings of the source over the last two decades, either as single band or coordinated multiwavelength observations, have revealed complex variability patterns and correlations between different energy bands.

Except for one orbit observation as part of the Guest Observer program on 19-21 November 2000 (Maraschi \et 2002), \pks\ has been repeatedly observed during other 8 orbits of \xmm\ as a calibration source. A detailed temporal analysis of the X-ray data from the first 6 orbits was presented in Edelson \et (2001) and  Zhang \et (2005, 2006; Paper I and II hereafter). Foschini \et (2006) presented average OM magnitudes, X-ray spectral properties and SEDs using part of these observations. In this paper we present a detailed analysis of the multiwavelength variability from optical to X-rays, using the simultaneous data obtained with OM and pn. The aim is to probe whether the X-ray variability properties of the source can be extended to the optical-UV bands, in order to test in detail the cooling and acceleration behaviour of the particle distribution responsible for the synchrotron emission. To our knowledge, the data presented here provide the most intensive simultaneous UV and X-ray monitoring achieved with \xmm\ for a blazar. 

In \S~\ref{sec:obs} we present the observations and data reduction; the multiwavelength variability analysis is described in \S~\ref{sec:om}. We compare our results with previous coordinated multiwavelength observations and discuss their physical implications in \S~\ref{sec:disc}.


\section{The XMM-Newton Observations and Data Reduction}\label{sec:obs}

During the observations of \pks\ over nine orbits, spanning over about 6 years, from 2000 May to 2005 November, all \xmm\ instruments, i.e., EPIC-pn, EPIC-MOS, RGS, and OM, functioned with various configurations. In this paper, we concentrate on OM and pn data from the first three orbits, i.e., orbits 087, 174 and 362, and present a detailed analysis of multiwavelength variability in the optical/UV and X-ray bands. The OM data from other orbits are not suitable for variability analysis because only one or two images are available for each filter, but they are useful to construct the SEDs with simultaneous X-ray data over a long period. We consider X-ray data, relative to the same orbits as the OM data, only from the pn camera because they are less affected by photon pile-up and have higher signal-noise-ratio with respect to the MOS cameras. All \xmm\ data were processed using the latest \xmm\ Science Analysis System (SAS) version 6.5 and the latest available calibration data. The journal of OM observations and related pn observations of the present analysis is shown in Table~\ref{tab:obs}, where an identification number is allocated to each observation.

The OM exposures were taken in standard imaging mode with different filters. We reduced OM imaging data using the standard \textit{omichain} pipeline. The OM images of orbits 087 and 174 were obtained through UV filters, either UVW2 (2120~\A) or UVW1 (2910~\A). The source counts were extracted in a circular region with $r=12$ pixels, and the background counts were subtracted from a concentric annular region between pixels 37 and 42. The optical U (3440~\A), B(4500~\A), V (5430~\A) filters were used for orbit 362 exposures; again we extracted the source counts in a circular region with $r=12$ pixels and subtracted the background counts from a concentric annular region between pixels 14 and 25.

For imaging mode of the pn exposures (orbits 087 and 174), the source counts were extracted from an annulus, centered on the peak of source counts, to avoid photon pileup effects. The inner and outer radii of the annulus, determined by using the SAS task {\it epatplot}, are 10'' and 40'', respectively. The part of the orbit 362 observation (362-1 as identified in Paper II) used here was obtained in timing mode, for which pileup can be neglected. We extracted the source photons from rows $25\ge $~RAWX~$\le 42$, centered on the brightest strip of the source. Moreover, we selected only single pixel events (pattern=0) with quality flag=0 to further minimize photon pileup effects for both the imaging and timing modes. The high particle background periods were examined by computing the hard ($E>10$~keV) count rates in full frame of the exposed CCD. We found that the background was low and stable for orbits 174-1 and 362-1, while a few background flares occurred in orbits 087-1, 087-2 and 174-2. As in the following analysis we mainly focus on the soft X-ray energy band, that is the least affected by high particle background, we do not cut out the relatively high background periods. We then extracted the background events from regions least affected by source photons to estimate the fraction of background vs source counts. In the 0.2-0.8~keV (imaging mode) and 0.6-0.8~keV (timing mode) bands the source count rates are of the order of $\sim 1000$ times the background ones, and even in the 2.4--10~keV band the background is about $5\%$ of the source count rates. Finally, the background count rates were subtracted.
        

\section{Results}\label{sec:om}

\subsection{Orbit 087}

\begin{figure}
\plottwo{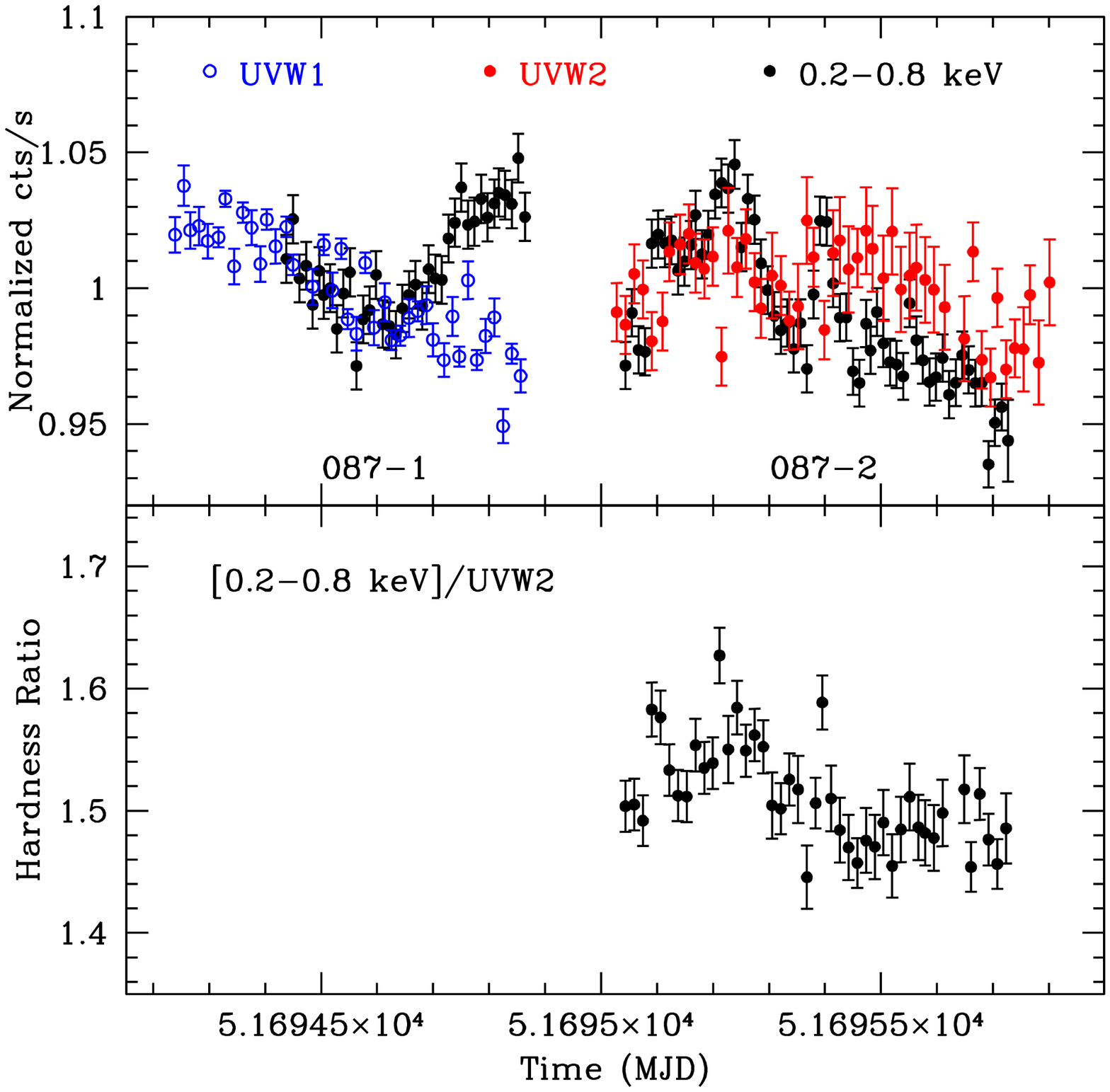}{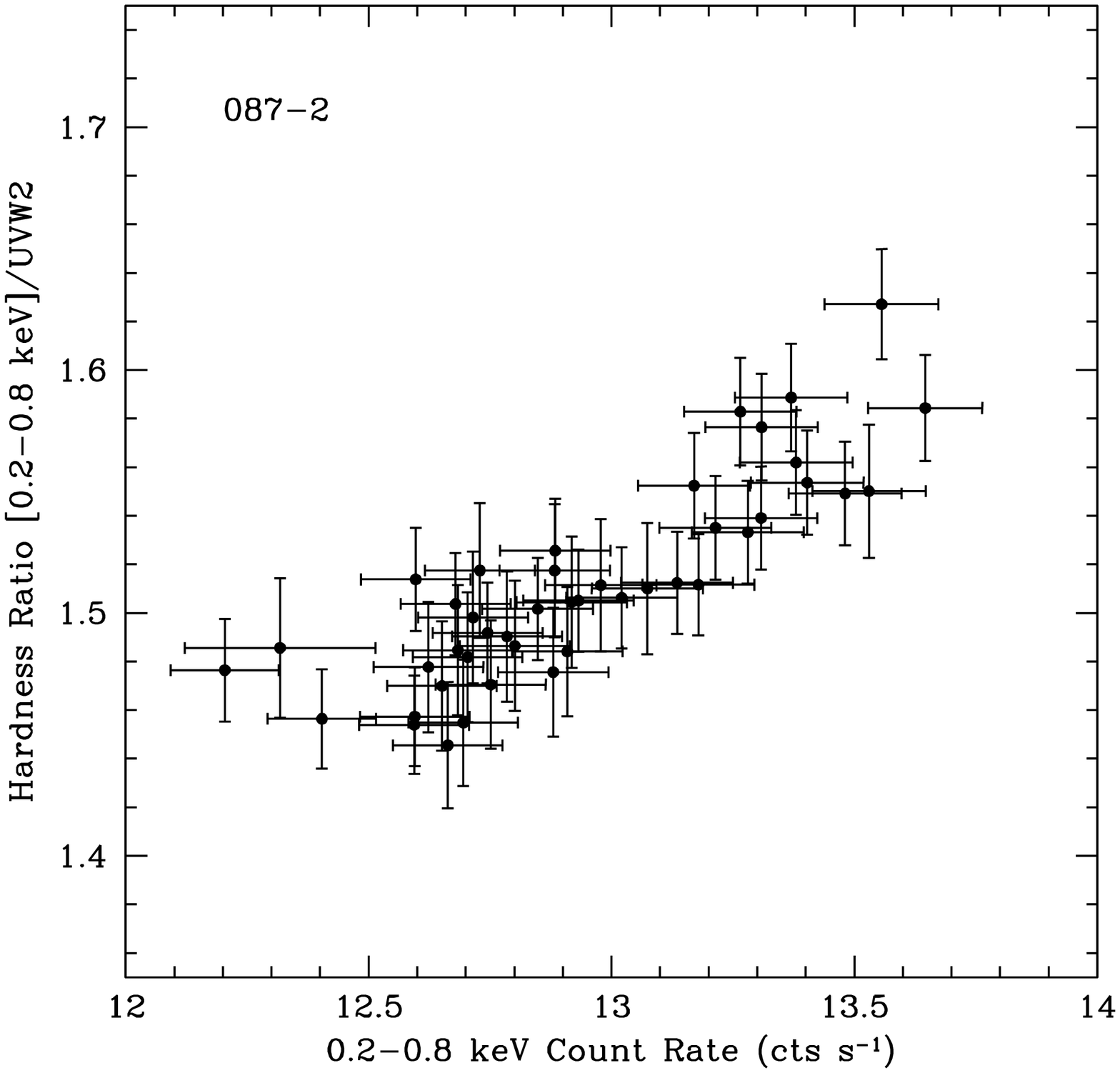}
\caption { \footnotesize 
Orbit 087 observation. 
\textit{Left}: The top panel plots the OM UVW1 (gray open circles), UVW2 (gray solid circles) and pn 0.2--0.8~keV (black) light curves normalized to their respective averages. The bottom panel plots the variations of the 0.2--0.8~keV/UVW2 hardness ratios with time for the 087-2 interval.  
\textit{Right}: The correlation between the hardness ratios and the 0.2--0.8~keV count rates for the 087-2 interval.  
Both the light curves and the hardness ratios are binned in 1000~s.} 
\label{fig:lchr:087}
\end{figure}

This observation can be split in two parts. For OM observations, the first part (087-1) consists of 40 images (exposures) obtained with the UVW1 filter, while in the second part (087-2) 50 images were obtained through the UVW2 filter. The exposure time of each image is 1000~s for both filters. This results in an almost continuous coverage of the UV light curve for 087-1 and 087-2, while readout time led to regular unexposed interval of $\sim$~325~s between any two neighboring points of each light curve. The top left panel of Figure~\ref{fig:lchr:087} shows the UVW1 (087-1, gray open circles) and the UVW2 (087-2, gray solid circles) light curves normalized to their respective averages. Given the small difference in wavelength, the normalized UVW1 and UVW2 light curves can be roughly viewed as a single light curve in the same energy band. Both UV light curves show visible variability, but have different behaviour. The 087-1 light curve presents a long decaying trend, while the 087-2 one is dominated by two flare-like events.

The pn observations were simultaneous to the OM exposures. Edelson \et (2001) presented a temporal analysis of the pn observations. They showed that the variability amplitude increased with increasing energy, and the variability between different energy bands correlated with no measurable lags. The correlations tend to become weaker with increasing energy differences, in agreement with the results of Paper II obtained with other pn observations. In order to compare the variability between the X-ray and UV bands, we extracted the pn light curves in three energy bands (see Paper II), i.e., 0.2--0.8~keV (soft), 0.8--2.4~keV (medium) and 2.4--10~keV (hard), and re-binned them over the 1000~s long UV exposure time. The soft 0.2--0.8~keV (black) light curve, normalized to its average, is also plotted in the top left panel of Figure~\ref{fig:lchr:087} for a direct comparison with the UV light curve.

The UVW1 light curve did not track the soft X-ray one during 087-1 interval. The X-ray fluxes increased significantly toward the end of this interval, while there was no evidence for increases of the UVW1 flux at the corresponding period. This is different from the good correlation between the light curves at different X-ray bands. However, the smaller variability and larger scatters in the UVW1 fluxes compared to the X-ray ones, and the lack of X-ray data at the start of this interval and the gap between intervals 087-1 and 087-2, prevent us from deriving any clear correlation between the UVW1 and X-ray light curves. Therefore, in the following we do not discuss this period of observations. 

\begin{figure}
\plottwo{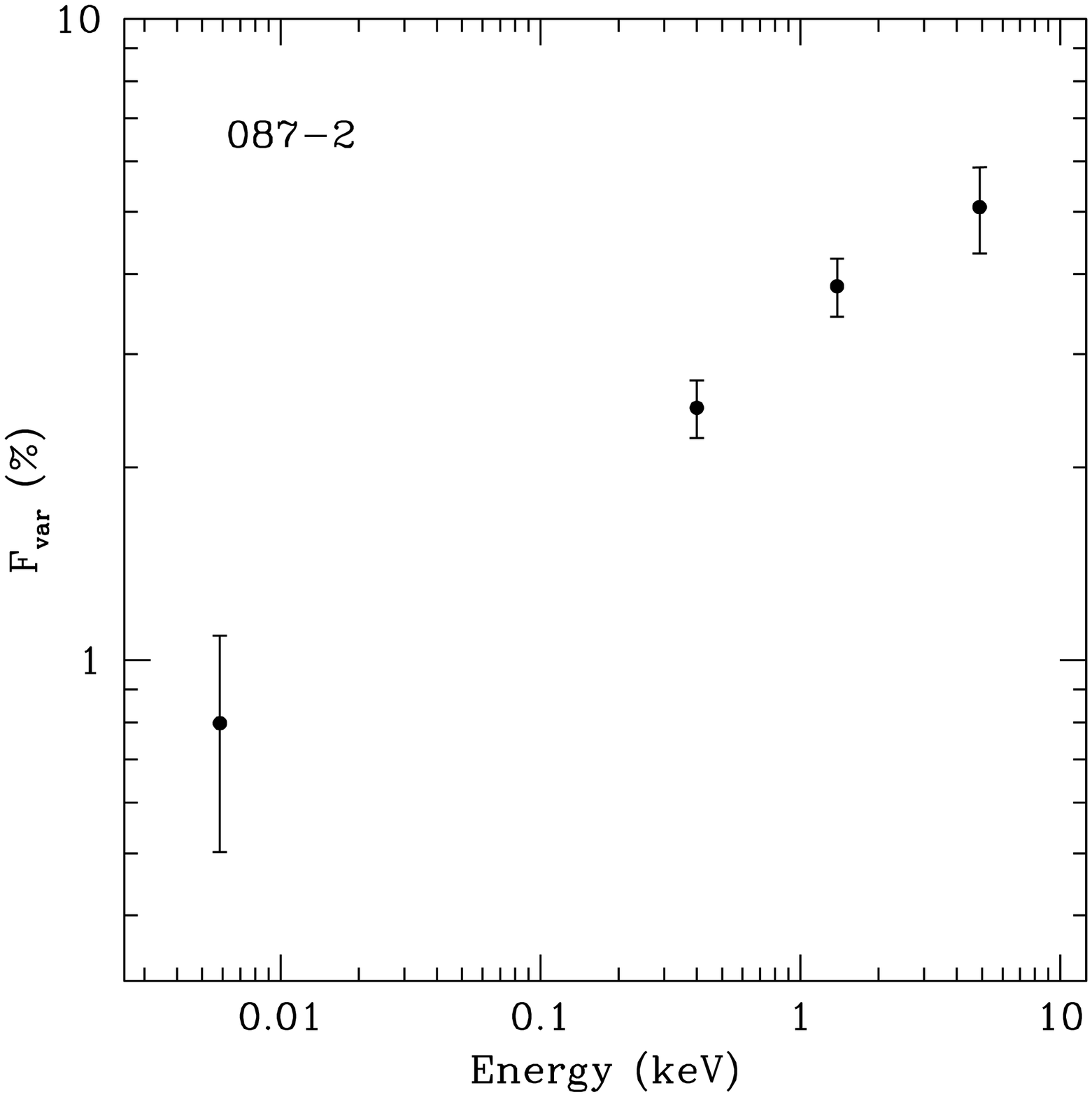}{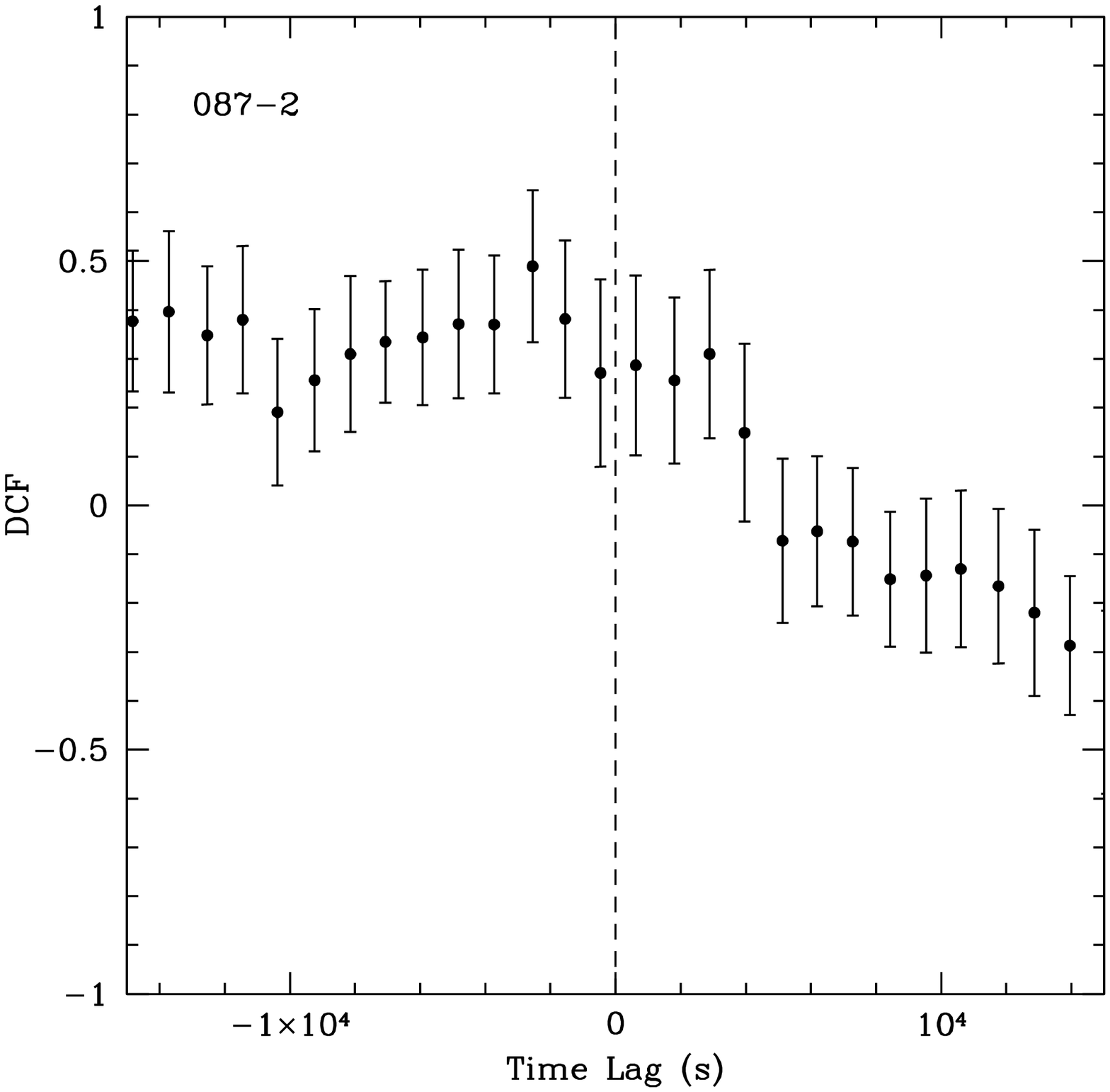}
\caption { \footnotesize Orbit 087-2 observation. 
\textit{Left}: The fractional rms variability amplitude as a function of energy.
\textit{Right}: The central $\pm 15$~ks range of the DCF between the UVW2 and 0.2--0.8~keV light curves. } 
\label{fig:rmsccf:087}
\end{figure}

During the interval 087-2, the UVW2 light curve appears to follow the soft X-ray one (top left panel of Figure~\ref{fig:lchr:087}). The left panel of Figure~\ref{fig:rmsccf:087} shows the energy dependence of the fractional $rms$ variability amplitude ($F_{\rm var}$) of the UVW2 and X-ray bands. The energy dependence of ($F_{\rm var}$) found in X-rays can be extrapolated to the UVW2 band: the UVW2 variability amplitude is substantially smaller than the X-ray one.

The bottom left panel of Figure~\ref{fig:lchr:087} shows the temporal evolution of the hardness ratio (HR) of soft X-ray to UVW2 band. As expected, the HR variations generally follow the soft X-ray flux variations given the lower UVW2 variability, indicating that the UV-X-ray spectral changes are mainly related to changes of X-ray intensities. The right panel of Figure~\ref{fig:lchr:087} reports the HRs as a function of the soft X-ray count rates. Indeed, the two quantities are well correlated -- the UV-X-ray spectrum becomes flatter when the X-ray flux is higher. Due to the smaller UVW2 variability, the HRs are indeed weakly correlated with the UVW2 fluxes.

Even though the UVW2 and X-ray light curves are simultaneous, they do not match exactly in time because the OM observing mode is different from the pn one (the latter is counting continuously while the former is not)\footnote{The HRs of the soft X-ray to UVW2 fluxes are obtained from the pn flux most closely matching in time the UVW2 one. In fact, the pn and UVW2 light curves can match exactly in time if one selects the pn exposure bins with the UVW2 exposure time intervals. This however reduces the pn observation.}. Therefore, we calculate the cross-correlation function (CCF) with the discrete correlation function (DCF) method (Edelson \& Krolik 1998). The right panel of Figure~\ref{fig:rmsccf:087} presents the central 15~ks part of the DCF between the UVW2 and the soft X-ray light curves for 087-2 (a positive lag indicates that the higher energy emission lags the lower energy one). The DCF is strongly asymmetric toward negative lags, indicating that the UVW2 variations may lag the soft X-ray ones. The lack of a clear peak in the DCF prevents us from quantifying any lag. 

The DCFs between the UVW2 and medium/hard X-ray bands are similar to that between the UVW2 and soft X-ray band for 087-2. We also calculated DCFs between different X-ray bands. The results are very similar to those presented by Edelson \et (2001), indicating that the X-ray variations at different energies are correlated with no measurable lags and the correlations tend to become weaker with increasing cross-correlated energy differences.

\subsection {Orbit 174}

\begin{figure}
\plottwo{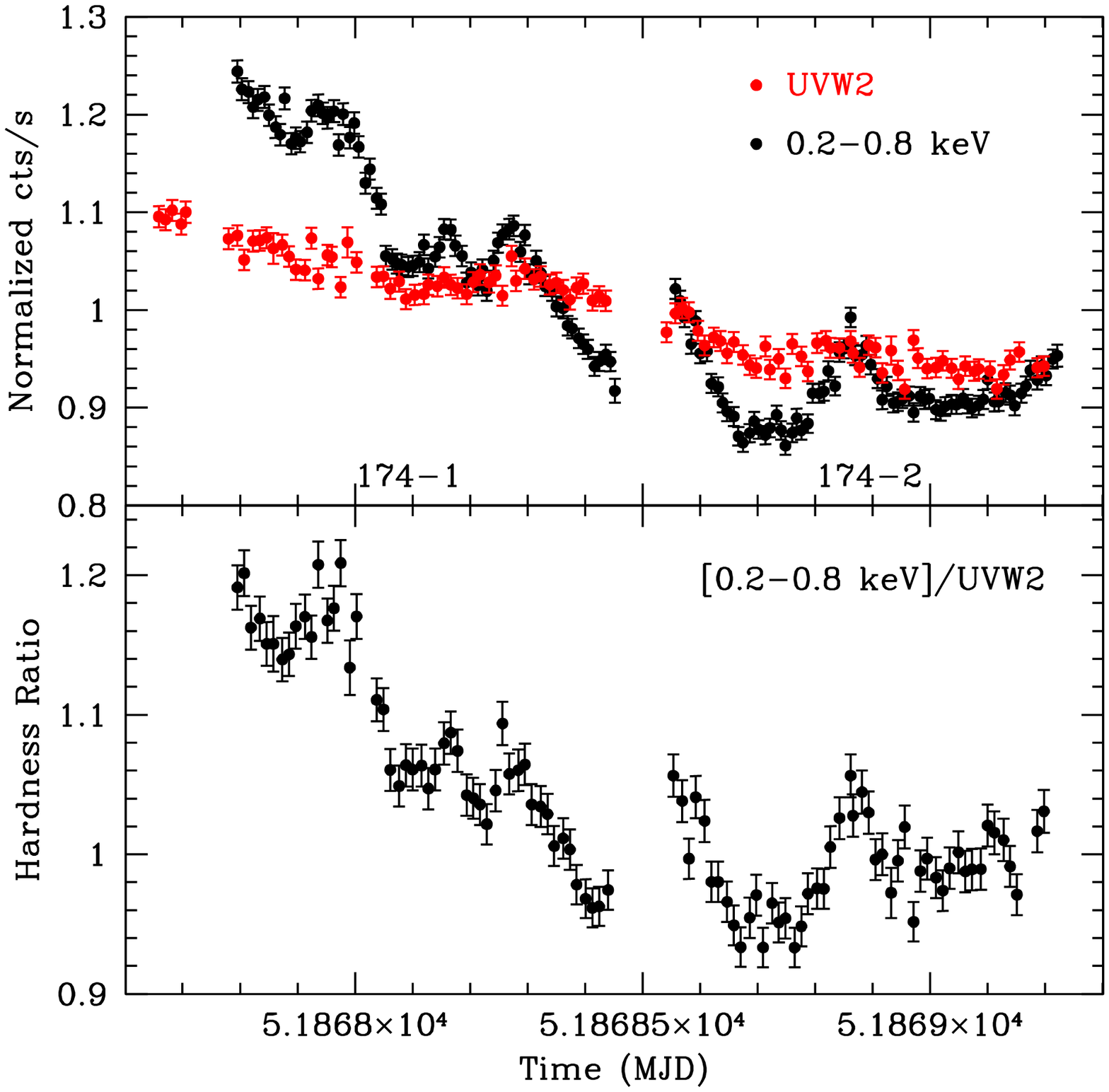}{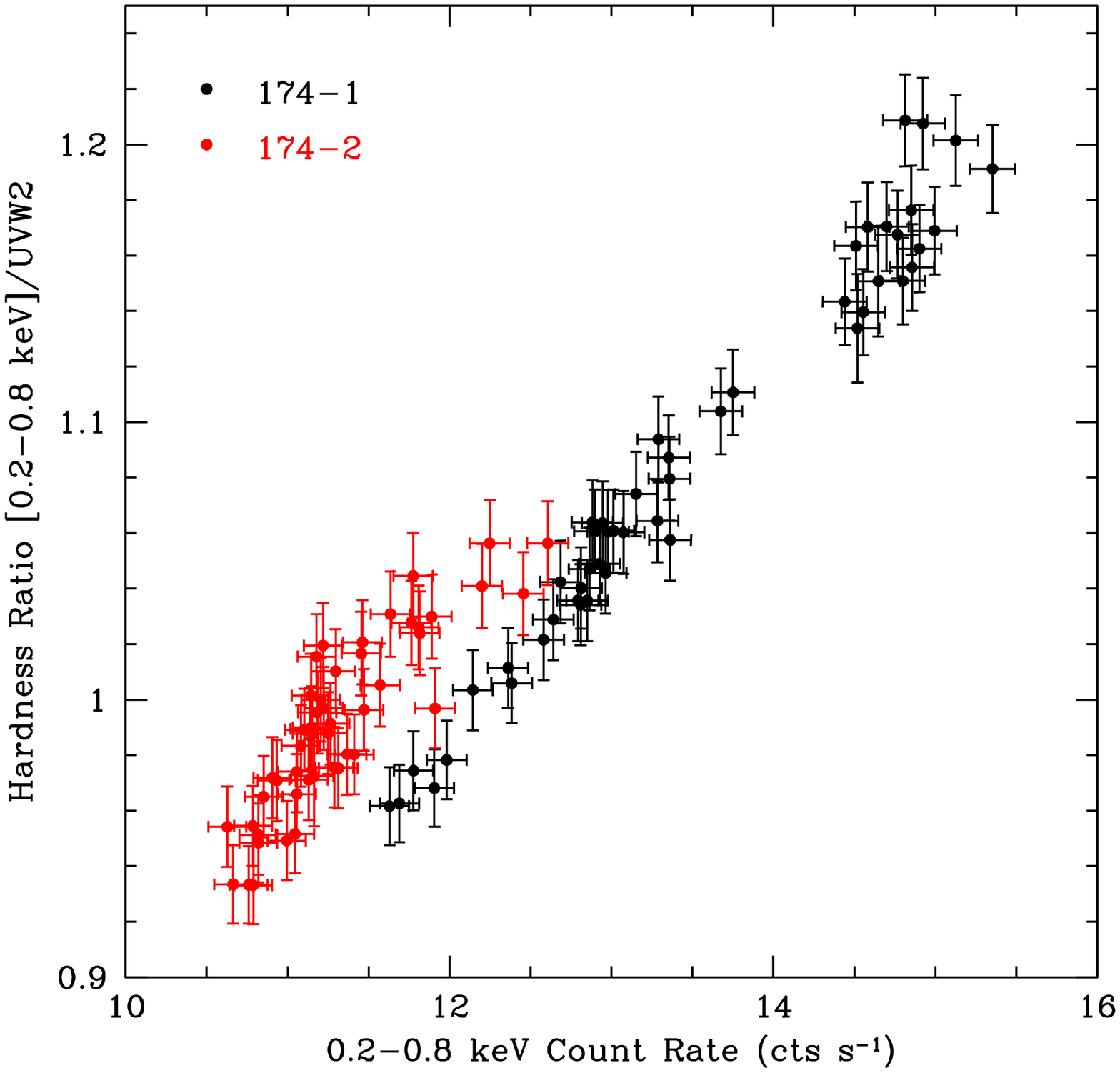}
\caption { \footnotesize 
Orbit 174 observation. 
\textit{Left}: The top panel plots the OM UVW2 (gray) and pn 0.2--0.8~keV (black) light curves normalized to their respective averages. The bottom panel shows the variations of the 0.2--0.8~keV/UVW2 hardness ratios with time. 
\textit{Right}: The correlation between the hardness ratios and the 0.2--0.8~keV count rates. Both the light curves and the hardness ratios are binned in 800~s.  } 
\label{fig:lchr:174}
\end{figure}

The OM images were obtained through the UVW2 filter, and the exposure time was 800~s for each image. The top left panel of Figure~\ref{fig:lchr:174} plots the UVW2 (gray) light curve normalized to its average. The OM observations consist of three sections. The first section (174-0) has 5 exposures without corresponding pn observations. The second (174-1) and the third (174-2) sections, simultaneously monitored with the pn, have 50 exposures each. This yields an almost continuous UVW2 light curve over $\sim 1.5$~days. Maraschi \et (2002) presented the simultaneous UVW2 and X-ray light curves. The UVW2 light curve is obviously variable, dominated by a long decaying trend with superimposed small amplitude flickers.

\begin{figure}
\plottwo{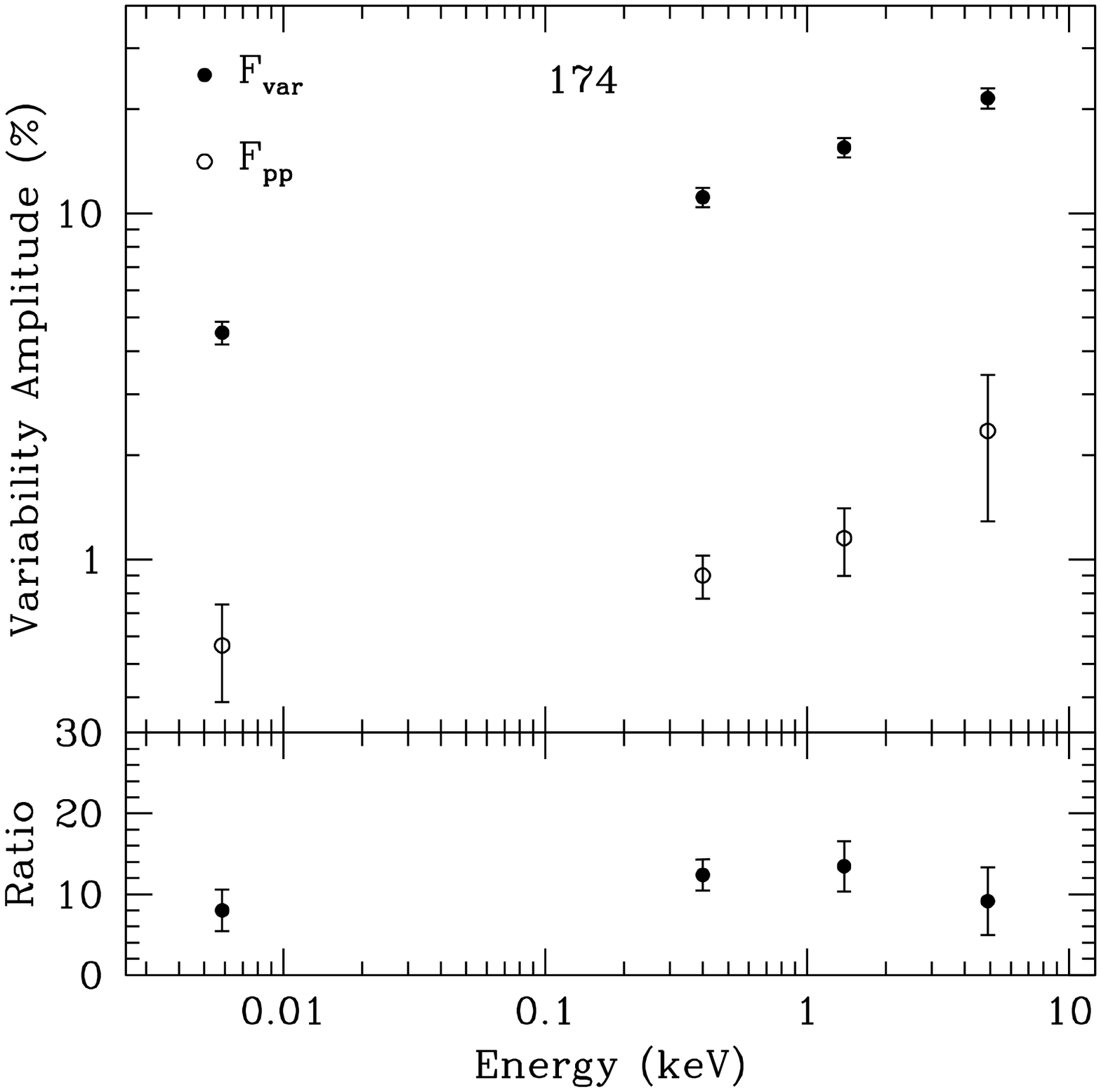}{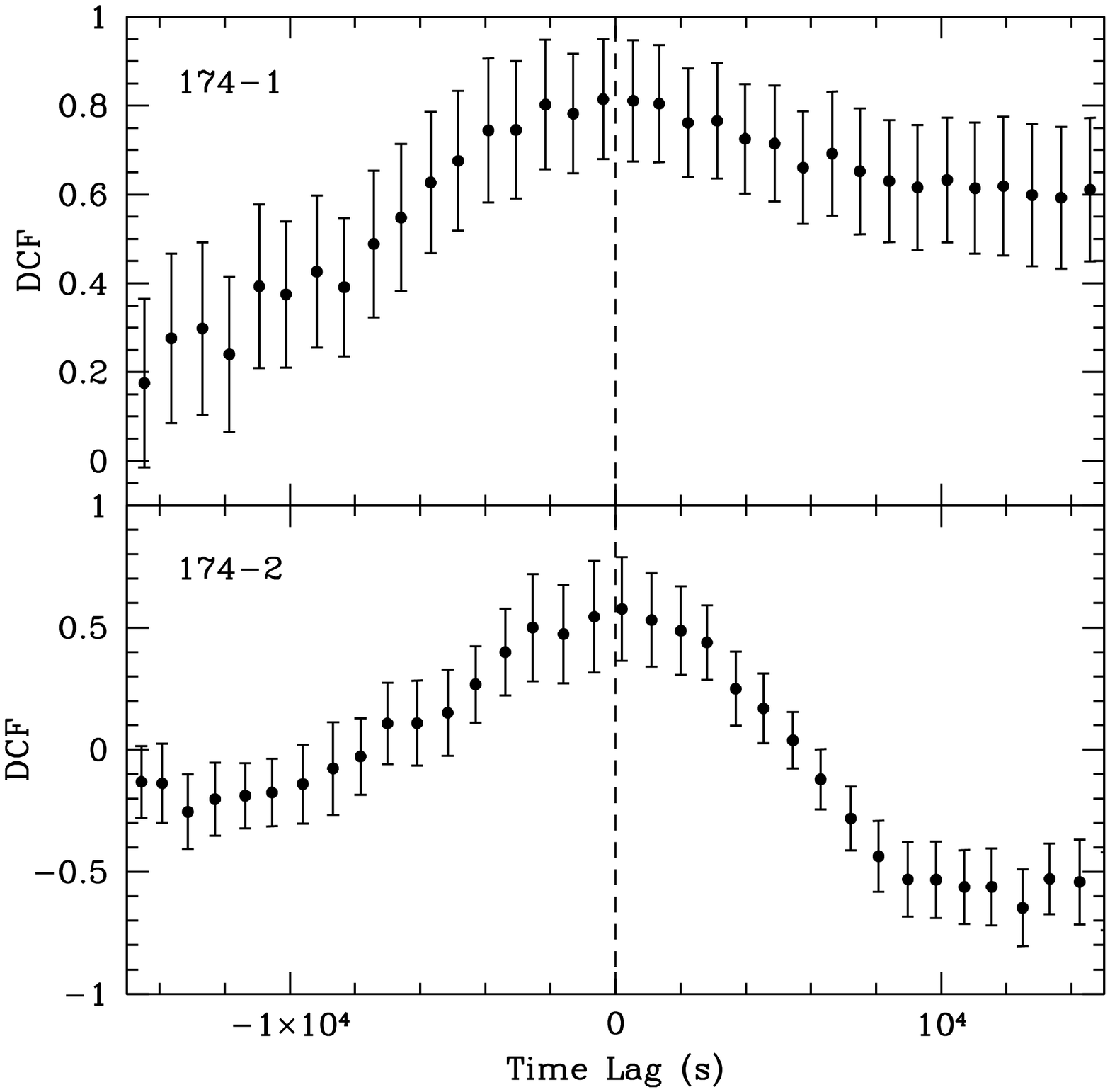}
\caption { \footnotesize Orbit 174 observation. 
\textit{Left}: The fractional rms variability amplitude as a function of energy over two different timescales (top panel): the upper \fvar\ points show the variability amplitude integrated over timescales between the total observational length and $\sim 800$~s, and is dominated by variations on timescale comparable to the observation duration because of red noise variability; the lower \fpp\ points indicate the rms deviation between adjacent points (i.e., point-to-point rms), sampling variability on short timescales ($\sim 800$~s). Bottom panel: the ratio of the two rms spectra, i.e., the ratio spectrum. 
\textit{Right}: The central $\pm 15$~ks range of the DCFs between the UVW2 and the 0.2--0.8~keV light curves. }  
\label{fig:rmsccf:174}
\end{figure}

The pn light curves (300~s binning) have been presented in Paper II. In order to compare the variability between the UVW2 and X-ray bands, we re-binned the pn light curves over the 800~s long UVW2 exposure time in three energy bands as above and in Paper II. The soft 0.2--0.8~keV (black) light curve, normalized to its average, is also shown in the top left panel of Figure~\ref{fig:lchr:174} for a direct comparison with the UVW2 one. The UVW2 light curve follows the X-ray one. However, the UVW2 variability amplitude is significantly smaller than the soft X-ray one, since the overall decrease in the UVW2 light curve is much flatter than that in the soft X-ray one. The fractional $rms$ variability amplitude ($F_{\rm var}$) is $4.5$\% and $10.8$\% for the UVW2 and soft X-ray band light curves, respectively. Paper I already showed that the X-ray variability amplitude is larger toward higher energy. The UVW2 variability amplitude is consistent with the extrapolation of the energy dependent X-ray variability amplitude. This is clearly seen from the top left panel of Figure~\ref{fig:rmsccf:174} (solid circles) that shows a very good correlation between $F_{\rm var}$ and photon energy. The top left panel of Figure~\ref{fig:rmsccf:174} (open circles) also shows that the point-to-point fractional $rms$ variability amplitude ($F_{\rm pp}$ , see Paper I for its definition) depends on photon energy as well. However, it appears that the ratio of $F_{\rm var}$ to $F_{\rm pp}$ (the bottom left panel of Figure~\ref{fig:rmsccf:174}) does not depend on energy, suggesting that the slopes of the power spectral density (PSD) are probably energy-independent. Therefore, the smaller amplitude of the UVW2 variability is connected to the smaller normalization of the PSD amplitude compared to the X-ray one (see also Paper I).

The bottom left panel of Figure~\ref{fig:lchr:174} shows the variations of the HRs of the soft X-ray to UVW2 fluxes. The HR variations generally follow the flux variations, especially in the soft X-ray band. This again indicates that the origin of the flux-related UV-X-ray spectral changes is related to the soft X-ray variability as expected from the small UVW2 variability. This behaviour is clearly shown in the right panel of Figure~\ref{fig:lchr:174}, which plots the HRs as a function of the soft X-ray count rates. We notice a deviation of the correlation defined by the interval 174-1 (at high count rates) and 174-2 (at low count rates), showing a harder spectrum for the 174-2 period. This might be interpreted if a less variable and harder emission component contributes more to the observed fluxes when the source is in a faint state.

As in the case of orbit 087, the UVW2 and X-ray light curves do not correspond to each other in time. We calculate the DCFs to examine the correlations between the UV and X-ray emission and determine any possible lags for the 174-1 and 174-2 exposures, respectively. The DCFs between the UVW2 and the soft X-ray light curves are plotted in the right panel of Figure~\ref{fig:rmsccf:174}. Both the DCFs peak around zero lag, and the maximum correlation (\rmax) is 0.81 and 0.58 for 174-1 and 174-2 DCF, respectively. This indicates that the UVW2 and soft X-ray variations are correlated, as already suggested by the normalized light curves. However, the two DCFs are not identical. The 174-1 DCF is asymmetric toward positive lags. This indicates that the soft X-ray variability may lag the UVW2 one over the long decaying trend, while they may be simultaneous over short timescales. The DCFs between the UVW2 and the medium/hard X-ray band are more asymmetric toward the positive lags. This behaviour is very similar to what obtained within the X-ray band, as presented in Paper II (left panel of Figure~6). The 174-2 DCF is slightly asymmetric toward negative lags, indicating that the UVW2 and the soft X-ray variations are almost simultaneous.
We quantify the possible lag and its significance using Monte Carlo simulations (Peterson et al. 1998). We calculated the probability that the lag is detected as either a negative (soft) or a positive (hard) lag. The results are listed in Table~\ref{tab:lag}. A probability larger than 95\% is considered as significant. Both the uncertainties and the probabilities inferred from the simulations show that the lags are consistent with zero for both 174-1 and 174-2 intervals. However, this result should be taken with caution. The UVW2 light curve of orbit 174 shows a descending trend but no features (peaks or dips) that could be ``anchored '' to similar features in the X-ray light curve to define a ``lag''. Therefore, the ``zero lag'', although formally present, could be induced by a chance coincidence of the two ``descending'' long term trends in the UVW2 and X-ray light curves.

\subsection{Orbit 362} 

\begin{figure}
\epsscale{0.5}
\plotone{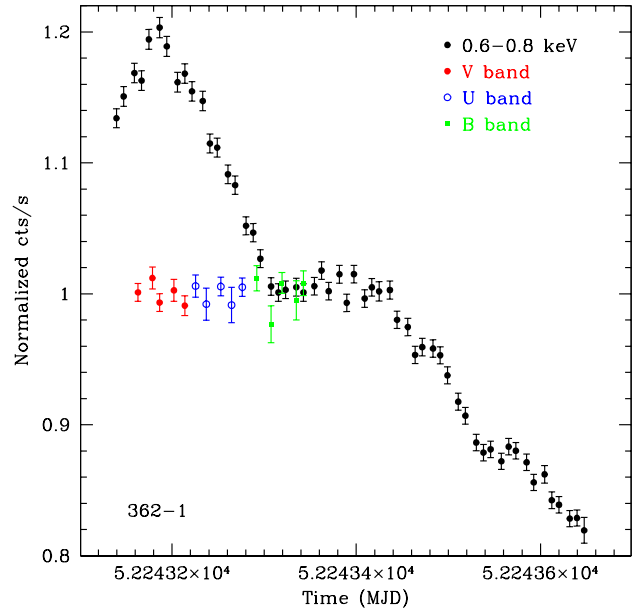}
\caption { \footnotesize Orbit 362 observation. 
 The OM V, U, B (gray) and pn 0.6--0.8~keV (black) light curves normalized to their respective averages. The light curves are binned over 800~s. }
\label{fig:lc:362}
\end{figure}

The OM observations were obtained with three different optical (V,B,U) filters. There are 5 continuous exposures for each filter in time sequence. The exposure time for each image is 800~s. Figure~\ref{fig:lc:362} reports, in time order, the V (gray solid circles), U (gray open circles) and B (gray solid squares) light curves normalized to their respective averages. The normalized V, U, and B light curves can be roughly viewed as a single continuous optical light curve given the limited wavelength differences. No pronounced variability is evident in the optical.

The time coverage of the optical light curves partly overlaps with the pn observation. Figure~\ref{fig:lc:362} also shows the normalized 0.6--0.8~keV pn light curve (black) binned over the 800~s long OM exposure time. While the optical light curves almost cover the period of the prominent X-ray flare, such flare was not detectable in the optical band.


\section{Discussion}\label{sec:disc}

In this work we presented a detailed multiwavelength timing analysis of \pks\ with the simultaneous optical-UV and X-ray data obtained during three orbits of observations with \xmm. To our best knowledge, these are the most intensive multiwavelength data gathered with different instruments onboard \xmm\ for a blazar. In the following we summarize and compare our main findings with those obtained from the previous multiwavelength observations of \pks, coordinated with different telescopes, which might be precious for the understanding of the complex variability behaviour of the source. 

\subsection{The \xmm\ Multiwavelength View}

The correlation between the UV and X-ray light curves of orbit 087 (2000 May 30-31) is not identical to that of seen about half a year later, during orbit 174 (2000 November 19-20). The light curves of orbit 087 appear to be dominated by multiple small amplitude flare-like events without a long term trend, while those of orbit 174 follow a long decaying trend, with superimposed short term timescale events on it. Moreover, it is worth noting that the orbit 174 observation may sample only a decaying phase of a possible major flare. During orbit 174 the UV variations closely follow the X-ray ones with no measurable lag, while during orbit 087 the UV emission weakly correlates with the X-ray one. This difference might be due to the smaller amplitude of variability in orbit 087 than in orbit 174, if indeed small amplitude variability is not correlated as strong as large amplitude variability events. Furthermore, the hardness ratio analysis shows that the UV-X-ray spectra during orbit 087-2 are significantly harder than those during orbit 174. Nevertheless in both cases the clear energy dependence of the variability amplitude found in the X-rays can be extrapolated to the UV wavelengths. This is consistent with what expected if both bands are dominated by a single synchrotron component. The hardness ratio analysis indicates that the UV-X-ray spectra harden with increasing X-ray brightness during both orbits, although with different amplitudes of spectral changes with fluxes. Finally, during orbit 362 (2001 November 30), the source did not show pronounced optical variability, while an intense X-ray flare (more than 40\% variability amplitude from minimum to maximum count rates) occurred in the same time interval as the optical observations.  

\subsection{Comparison with Previous Multiwavelength Observations} 

The \xmm\ multiwavelength observations of \pks\ revealed that the UV variability may correlate with the X-ray variability with no measurable lags in the two epochs separated by about half a year. This is different from the complex behaviour of the correlations between the UV and X-ray variations obtained from the two previous extensive multiwavelength observations of the source in 1991 November and 1994 May, coordinated with IUE, ROSAT, ASCA and ground-based telescopes. During $\sim 4.5$~days of observations in 1991 November, the source showed highly correlated and achromatic variations of small amplitude at optical, UV and X-ray wavelengths, characterized by multiple flares of roughly the same amplitude and duration. The X-ray variations probably led the UV ones by a couple of hours (Edelson \et 1995). In contrast, during $\sim 10$~days of the UV observations in 1994 May with IUE and EUVE, and $\sim 2$~days of the intense X-ray observations with ASCA (this campaign also involved observations with ROSAT and ground-based radio, infrared, and optical telescopes), the source showed a well-defined flare whose properties strongly depend on photon energy. With increasing wavelengths, the variability amplitude significantly decreased, and the temporal profile of the flare broadened. The variations at different wavelengths were strongly correlated at significant lags, and the lags became larger with larger differences in wavelengths. Specifically, the UV variations lagged the X-ray ones by $\sim 2$~days (Urry \et 1997).

The first simultaneous multiwavelength campaign involving very high energy (VHE) $\gamma$-ray (with H.E.S.S.) were performed in 2003 October and November (Aharonian \et 2005b). The observations showed that the source was variable during a low state. However, the VHE $\gamma$-ray variations were clearly more intense than those in the X-ray and optical bands, and no correlations between them were detected over the timescales of the observations. The \xmm\ orbit 362 observation may repeat a similar phenomenology on short timescales: a well resolved major X-ray flare was not followed by a corresponding optical flare.

The \xmm\ observations are superior to the other multiwavelength observations. (1) Because IUE was a geostationary satellite with 96 min time resolution required to get a UV spectrum, and ROSAT and ASCA were low Earth-orbit satellites ($\sim 1.6$~hours period), the observations with them were characterized by low time resolution and/or periodical gaps. \xmm\ is a high Earth-orbit satellite ($\sim 2$~days period), so the observations with it can be almost continuous. This implies that the temporal resolution of OM is much higher than that of IUE: the typical binning time is $\sim 15$~min and $\sim 96$~min for OM and IUE, respectively; (2) the signal-noise ratio of the pn X-ray observations is much higher than those of ASCA and ROSAT. However, the observation length is $\sim 1$~day for \xmm\ and several days for IUE/ASCA. Therefore, \xmm\ mostly characterized the intra-day multiwavelength variability with the highest temporal resolution, while the other coordinated UV-X-ray observations mainly emphasized extensive inter-day multiwavelength variations of the source. If the 2-day delay of UV vs X-rays were present during the \xmm\ observations, it would not have shown up because those were much shorter than 2 days. However, it is also possible that the UV-X-ray correlation and lag may change with time.    
 
\subsection{Physical Implications}

The multiwavelength observations obtained so far showed that the UV and X-ray emission of \pks\ have consistent variability properties, i.e., the systematic dependence of the variability amplitude on energy, from UV to X-rays, and the similar trends of the UV and X-ray light curves. These similarities/consistences imply a common origin, plausibly synchrotron radiation, for the emission in these bands. However, these observations also showed different patterns of multiwavelength variations, namely changes of UV-X-ray lags and detailed differences of the UV-X-ray variability on the short timescales. These differences indicate that the constraints on the radiation models would be different at the different energy bands from epoch to epoch, requiring, e.g., changes in the parameters characterizing the emitting region, different mechanisms operating at different time, and/or different dynamical timescales in the UV and X-ray bands.

In the simplistic homogeneous scenario, the synchrotron emission is initiated by a single injection of high energy electrons in a magnetized volume, followed by energy-dependent cooling: the higher the energy, the faster the cooling. An instantaneous acceleration of high energy electrons is a representation of such injection. The observed variability properties of the source will depend on the details of the acceleration (e.g., Katarzi\'nsky \et 2005), and the energy that is first emitted will depend on the balance between the acceleration and cooling timescales of the relativistic electrons. Within such scenario attempts have been made to interpret the remarkably different inter-band (especially X-ray) time lags detected at different epochs (e.g., Kirk, Rieger, \& Mastichiadis 1998). The energy-dependent amplitude of variability is a natural result of the energy-dependent cooling. When cooling is responsible for the variability behaviour, the observed soft lags, i.e., the low energy variations lagging the high energy ones, provides a way to estimate the physical parameters of the emitting region (e.g., Zhang \et 2002)\footnote{In the following we concentrate on soft lags. A discussion on hard lags (high energy emission lagging the low energy one) can be found in our previous work (Zhang 2002; Zhang \et 2002). In Paper II, we also point out that the probability of detecting a soft lag is higher than detecting a hard lag, as inferred from previous results. This may indicate that the acceleration timescales is usually much shorter than the cooling timescales.}. In the observer's frame, we have
\begin{equation}
B\delta^{1/3} =209.91 \times \left (\frac{1+z}{E_{\rm l}}\right)^{1/3}
	\left [\frac{1 - (E_{\rm l}/E_{\rm h})^{1/2}}
        {\tau_{\rm soft}} \right ]^{2/3} \quad {\rm Gauss}  \,,
\label{eq:soft}
\end{equation} 
where $z$ is the source's redshift, $B$ the magnetic field (in Gauss) and $\delta$ the bulk Doppler factor of the emitting region, and $\tau_{\rm soft}$ the observed soft lags (in second) between the low ($E_{\rm l}$) and high ($E_{\rm h}$) energy (in keV), respectively.
Equation~(\ref{eq:soft}) implies that the larger the lag, the smaller the combination of $B$ and $\delta$. If, as often assumed, $\delta$~($\sim 10$) does not change significantly with time, the lag is closely related to $B$ as $B \propto \tau_{\rm soft}^{-2/3}$. In order to be confident on whether $B$ does change significantly with time, one needs proper measurements of the lags with observations obtained at different epochs. However, as we stated in the last section, the UV and X-ray light curves obtained with \xmm\ are not comparable with those obtained during previous coordinated multiwavelength observations. In particular, the \xmm\ orbit 174 observations may sample only the decaying phase of a major flare while the 1994 May campaign caught a complete flare. The UV-X-ray lags measured from these two campaigns may thus have different meanings. Therefore, it is still premature to conclude that the significantly different UV-X-ray lags measured between the two sets of observations indicate intrinsic changes of $B$.

Furthermore the interpretation of the observed lags may be very likely affected by, e.g., inhomogeneous emitting region(s), such as in a stratified shock model or in an energy-dependent volume. The achromatic multiwavelength variability showing a quasi-periodicity in 1991 November, at odds with the predictions of simple synchrotron emission models, could have been even caused by micro-lensing by stars in an intervening galaxy (Treves \et 1997) or associated with helical trajectories of moving knots in a relativistic jet (Urry \et 1993; 1997).

All the results appear to indicate that the variability behaviour observed so far does not provide robust constraints to test or pin down the physical properties of the emitting region. Well-defined major flares, possibly produced by a single variability episode, might still provide the most favourable situation to probe, e.g., any connection between the sign of the lag and the rise and decay of the flux, and any relation between the lag (sign) and the peak energy of the synchrotron emission. 




\acknowledgments
We thank the anonymous referee for constructive suggestions and useful comments. 
This research is based on observations obtained with \xmm, an ESA science mission with instruments and contributions directly funded by ESA Member States and NASA. YHZ acknowledges Project 10473006 supported by National Natural Science Foundation of China and the Key Project of Chinese Ministry of Education (NO 106009). JMB acknowledges Project 10573030 supported by National Natural Science Foundation of China. AC acknowledges the Italian MIUR for financial support.


\clearpage
\begin{deluxetable}{ccrllllll}
\tabletypesize{\footnotesize}
\tablecolumns{8}
\tablewidth{0pc}
\tablecaption{Observational Journal of \pks\ with \xmm\ }
\tablehead{
\colhead{Orbit} &\colhead{Observation Id\tablenotemark{a}} &\colhead{Date} &\colhead{Detector\tablenotemark{b}} &\colhead{Filter} &\colhead{Duration} &\colhead{Exposure} &\colhead{CR\tablenotemark{c}}\\
&  &\colhead{(TT)} &  & &\colhead{(10ks)} &\colhead{(10ks)} & 
}
\startdata
087 &0124930101(087-1) &2000-05-30T05:33:33-20:39:17  &OM &UVW1   &40\tablenotemark{e}  &0.1\tablenotemark{f}  &72.7 \\
     &                  &           10:20:09-20:53:29 &pn SW    &Medium &3.80 &2.66 &13.2 \\
     &0124930201(087-2) &2000-05-31T00:34:43-19:20:43 &OM &UVW2   &50\tablenotemark{e} &0.1\tablenotemark{f}   &8.5 \\
     &                  &           00:52:59-17:21:38 &pn SW    &Medium &5.93 &4.16 &12.9 \\
174 &0080940401(174-0) &2000-11-19T15:40:50-17:08:57  &OM &UVW2   &5\tablenotemark{e}  &0.08\tablenotemark{f}    &13.1 \\
     &0080940101(174-1) &           18:42:13-10:41:33\tablenotemark{d} &OM &UVW2   &50\tablenotemark{e} &0.08\tablenotemark{f}    &12.4 \\
     &                  &           19:00:40-10:55:39\tablenotemark{d} &pn SW    &Thin  &5.73 &4.02 &13.5 \\
     &0080940301(174-2) &2000-11-20T12:56:53-04:56:12\tablenotemark{d} &OM &UVW2 &50\tablenotemark{e} &0.08\tablenotemark{f}   &11.4 \\
     &                  &           13:15:19-05:25:19\tablenotemark{d} &pn SW    &Thin  &5.82 &4.08 &11.3 \\
362 &0124930301(362-1) &2001-11-30T03:48:29-05:16:36 &OM    &V  &5\tablenotemark{e} &0.08\tablenotemark{f} &98.4 \\
     &                  &          05:21:58-06:50:04 &OM    &U  &5\tablenotemark{e} &0.08\tablenotemark{f} &223.3 \\
     &                  &          06:55:24-08:23:33 &OM    &B  &5\tablenotemark{e} &0.08\tablenotemark{f} &231.2 \\
     &                  &          03:12:05-15:30:29 &pn TI &Medium &4.32 &4.27 &28.3 \\

\enddata
\tablenotetext{a}{The number in parentheses is the identification number used in the text for the observation.}
\tablenotetext{b}{OM was operated in the standard imaging mode. For pn, SW indicates imaging small window, and TI timing mode.} 
\tablenotetext{c}{Background subtracted source count rate. The energy band is 0.2--0.8~keV for pn imaging mode, and 0.6--0.8~keV for pn timing mode.}
\tablenotetext{d}{Next day.}
\tablenotetext{e}{The total number of OM images.}
\tablenotetext{f}{The exposure time for each OM image.}
\label{tab:obs}
\end{deluxetable}
\begin{deluxetable}{lcccccc}
\tablecolumns{7}
\tabletypesize{\footnotesize}
\tablewidth{0pt}
\tablecaption{Time lags and Probability}
\tablehead{
\colhead{} &
           &\multicolumn{2}{c}{Lag (s)\tablenotemark{a}}
	&   &\multicolumn{2}{c}{Probability (\%)\tablenotemark{b}}\\
\cline{3-4} \cline{6-7}
\colhead{Observations} 
	&\colhead{$r_{\rm max}$} 
	&\colhead{$\tau_{\rm peak}$} &\colhead{$\tau_{\rm cent}$}
      & &\colhead{$\tau_{\rm peak}$} &\colhead{$\tau_{\rm cent}$}
}
\startdata
%
174-1 &0.81 &$542^{+797}_{-2696}$  &$119^{+438}_{-468}$ &&57.7 &71.7  \\
174-2 &0.58 &$194^{+916}_{-1791}$  &$85^{+812}_{-834}$ &&69.8 &53.6  \\
\enddata
\tablenotetext{a}{The lags are measured with peak (\tpeak) and centroid (\tcent) method (see Paper II). The quoted values are the medians of the simulations, and the errors are $68\%$ confidence range with respect to the medians.}
\tablenotetext{b}{The probability that the lag is detected as a hard lag.}
\label{tab:lag}
\end{deluxetable}

\clearpage

\end{document}